\documentclass[twocolumn,prl,preprintnumbers]{revtex4}
\usepackage{graphicx}
\usepackage{verbatim}

\begin{document}

\preprint{MIT-CTP/4147}

\title{Correlated Topological Insulators and the Fractional Magnetoelectric Effect}
\author{B. Swingle, M. Barkeshli, J. McGreevy, and T. Senthil}
\affiliation{Department of Physics, Massachusetts Institute of Technology,
Cambridge, MA 02139, USA }
\begin{abstract}
Topological insulators are characterized by the presence of gapless surface modes protected by time-reversal symmetry. In three space dimensions the magnetoelectric response is described in terms of a bulk $\theta$ term for the electromagnetic field. Here we construct theoretical examples of such phases that cannot be smoothly connected to any band insulator. Such correlated topological insulators admit the possibility of fractional magnetoelectric response described by fractional $\theta/\pi$.   We show that fractional $\theta/\pi$ is only possible in a gapped time reversal invariant system of bosons or fermions if the system also has {\em deconfined}  fractional excitations and associated degenerate ground states on topologically non-trivial spaces. We illustrate this result with a concrete example of a time reversal symmetric topological insulator of correlated {\em bosons} with $\theta =\frac{\pi}{4}$.  Extensions to electronic fractional topological insulators are briefly described.
\end{abstract}
\maketitle

Topological insulators are insulating phases of matter that enjoy gapless surface modes protected by time reversal invariance \cite{tirev1,tirev2}. For non-interacting electrons, these states can be characterized in terms of the non-trivial topology of their band structure \cite{tiband1,tiband2,tiband3}. Since the original theoretical proposals, experimental evidence supporting the existence of such states has accumulated in a number of materials \cite{hgte1,hgte2,tirev1,tirev2}. At the current theoretical frontier is the extension of these phenomena to interacting systems, where a characterization in terms of band structure is insufficient \cite{topomott,fqsh2}. To that end a useful alternative characterization (for $3D$ materials) in terms of the response to an external electromagnetic field has been proposed that makes use of the $\theta$ term $\frac{\theta}{2\pi}\frac{e^2}{2 \pi} \vec{E}\cdot \vec{B}$ \cite{tft1,tft2}. Non-interacting fermionic topological insulators have $\theta = \pi \text{ mod } 2\pi$ while trivial insulators have $\theta = 0 \text{ mod } 2\pi$.

In this paper we study time reversal symmetric topological insulating phases that cannot be smoothly connected to any band insulator.  We focus specifically on two questions.  First, is there a fractional generalization of the non-interacting topological insulator characterized by fractional $\frac{\theta}{\pi}$ while preserving time reversal symmetry?  This question was raised very recently in an interesting paper \cite{fti1}, although the details of the example suggested in that work are problematic, as we discuss below. Second, can a system of repulsively interacting bosons form a time reversal symmetric topological insulator with a non-zero $\theta$? As bosons cannot form a band insulator, such a phase, if it exists, is necessarily stabilized by interaction effects.   A positive answer in either case implies the existence of time reversal protected gapless surface states at an interface with a region with $\theta = 0$. Indeed, these gapless degrees of freedom are needed to `cancel' the naive time reversal non-invariance of a spatially varying $\theta$ in the interface region.

In this paper we will answer both questions. We first show quite generally that a time reversal symmetric fractional-$\theta/\pi$ topological insulator that is gapped in the bulk necessarily has {\em deconfined} fractionalized excitations. The presence of such excitations is signaled by the existence of a kind of topological order familiar from previous work on gapped fractionalized phases in two or more dimensions \cite{wen}. One consequence is non-trivial ground state degeneracy on topologically non-trivial spaces. We illustrate this by constructing an example of a fractional topological insulator of {\em bosons} which has $\theta = \frac{\pi}{4}$. This phase has fractionalized charge $1/2$ excitations with a bulk gap and string-like vortex excitations. These excitations are described in terms of a deconfined $Z_2$ gauge theory in $3+1$ dimensions. The system also has degenerate ground states on a closed topologically non-trivial space and time reversal protected gapless surface states.

Our construction is readily generalized to describe fractional {\em electronic} topological insulators. Furthermore, the bosonic topological insulator can be simply reinterpreted as a construction of a time reversal symmetric topological spin insulator of quantum magnets which conserve one component of the spin. We will briefly mention these generalizations toward the end of the paper.

We begin by considering a time reversal invariant insulator of bosons or fermions in $3+1$ dimensions with a gap for all excitations.  We further assume that the low energy theory contains a non-zero $\theta$ term in the response to an external electromagnetic field that couples to the conserved particle number.  With these assumptions we will show that a unique ground state on the three dimensional torus $T_3$ implies $\theta = \pi$.

The $\theta$ term takes the form
\begin{equation}
S_{\theta} = \frac{\theta}{2 \pi}\frac{e^2}{2 \pi} \int\, d^3 x \, dt\, \vec{E}\cdot \vec{B}
= \frac{\theta}{2 \pi}\frac{e^2}{4 \pi} \int F \wedge F.
\end{equation}
The normalization is chosen so that $\theta$ is $2 \pi $ periodic. There will also be a Maxwell-like term in the effective action, but this term is
time reversal symmetric, and only the $\theta$ term is potentially dangerous to time reversal symmetry.  The ground state to ground state amplitude is well defined for adiabatic processes because the ground state is unique and the gap is finite; it is given by
\begin{equation}
Z[\vec{E},\vec{B}] = C \exp{\left( i S_{\theta}[\vec{E},\vec{B}]\right)}.
\end{equation}

Consider a field configuration consisting of a background magnetic field in the $z$ direction which is uniform in the $xy$ plane.  The flux through
the non-contractible $xy$ two-torus is quantized \begin{equation}
\int_{xy} e B_z = 2 \pi n,
\end{equation}
and we will consider the minimal flux of $2 \pi / e$.  This flux is an allowed low energy configuration if the low energy excitations carry only integer charge. Note that we assume the insulator's unique gapped ground state persists in the presence of an infinitesimal field.  This assumption is physically reasonable if we allow arbitrary time reversal invariant perturbations. In the presence of this background magnetic field we insert the same minimal flux $2 \pi /e$ through the non-contractible $z$ loop of the three torus. Although the $\theta$ term is locally a total derivative, it contributes to bulk processes involving topologically non-trivial background field configurations.  The ground state to ground state amplitude for this process is $Z = C \exp{(i \theta )}$.
% two changes here

Now this system is time reversal invariant by assumption, so the response of the system to the time reversed configuration of electric and magnetic fields
must be the same.  The time reverse of the flux insertion process we considered still inserts $2 \pi /e$ flux, but the background magnetic flux changes sign to $- 2 \pi/ e$.  The ground state to ground state amplitude for this process is $Z = C \exp{(-i \theta)}$.  Thus the responses to these time reversed processes are only equal if $\theta = \pi$.  This proves our claim that a bulk gap, time reversal symmetry, and a unique ground state on $T_3$ are only consistent with $\theta = \pi$ or $\theta = 0$.

Thus to have a fractional $\theta$ angle in a gapped system, we must either break time reversal or have ground state degeneracy on $T_3$ and
other topologically non-trivial spaces.  The latter implies the presence of topological order of a kind familiar from the fractional quantum Hall effect and other fractionalization phenomena in space dimensions higher than one \cite{wen}. It goes hand in hand with the presence of {\em deconfined} fractional excitations in the bulk. We will now give a detailed construction of such a phase in a correlated bosonic system.

We consider hard-core bosons hopping in $3+1$ dimensions on a diamond lattice with two sites per unit cell. The bosons are taken to be at a commensurate density of one boson per unit cell.  Our goal is a time reversal invariant fractionalized phase of the bosons where the $U(1)$ boson number symmetry also possesses a non-zero $\theta$ angle for a background $U(1)$ gauge field. The normalization of the charge is fixed by requiring the boson operator $b_r$ to carry charge $1$ under the $U(1)$.  We employ a slave particle representation and construct a stable mean field theory for a topological insulator phase with the desired properties.  Write this boson as
\begin{equation}
b_r = d_{r 1} d_{r 2} = \frac{1}{2} \epsilon_{\alpha \beta} d_{r \alpha} d_{r \beta}
\end{equation}
where the two fermions $d_{r \alpha}$ carry charge $1/2$ under the boson number symmetry.  $\alpha, \beta = 1,2$ is a pseudospin index so that the $d_\alpha$ transform as a spinor under pseudospin $SU(2)$ rotations. We may thus view the boson as a pseudospin-singlet `Cooper pair' of these $d$ fermions.

As usual in the slave particle approach, this decomposition of the boson is redundant.  Any pseudospin $SU(2)$ rotation $d_r \rightarrow U_r d_r$ where $d_r = (d_{r 1}\, d_{r 2})^T$ and $U_r$ is an arbitrary $SU(2)$ matrix leaves the boson operator invariant. Because of this local redundancy, any low energy description involving the fractionalized ``slave'' particles $d_{r \alpha}$ must necessarily include gauge fields.  Below we will construct a stable mean field theory for a fractionalized phase which breaks the $SU(2)$ gauge structure down to $Z_2$.  Then the true low energy theory of the resulting phase will be a $Z_2$ gauge theory.

We assume a mean-field state for the fermions $d_{r \alpha}$ in which they form a topological band insulator with the $\alpha$ index playing the role usually played by physical spin. For concreteness we consider the tightbinding Hamiltonian on the diamond lattice introduced by Fu, Kane, and Mele \cite{tiband2} for a strong topological insulator.
\begin{equation}
\label{mfHam}
H = \sum_{\langle r r' \rangle} t_{r r'} d^\dagger_{r \alpha} d_{r' \alpha} +
i \sum_{\langle \langle r r' \rangle \rangle} \lambda^a_{r r'} d^\dagger_{r \alpha} \tau^a_{\alpha \beta} d_{r' \beta} + \text{h.c.},
\end{equation}
where $\langle r r' \rangle$ runs over nearest neighbors and $\langle \langle r r' \rangle \rangle$ runs over next nearest neighbors on the diamond lattice. The ``spin-orbit" interaction is defined by $\lambda^a_{r r'} = \lambda \epsilon^{a b c} n^{(1) b}_{r r'} n^{(2) c}_{r r'}$ where $n^{(1) a}_{r r'}$ and $n^{(2) a}_{r r'}$ are the nearest neighbor bond vectors, with $a$ regarded as a spatial index, traversed when hopping from $r$ to the next nearest neighbor $r'$.  As shown in \cite{tiband2}, by perturbing the nearest neighbor hopping $t_{r r'}$ away from uniform hopping we may enter a strong topological insulator phase. As the mean field Hamiltonian does not conserve any component of the spin, it follows that that the $SU(2)$ gauge structure implied by the slave particle representation is
broken down. Indeed only a $Z_2$ subgroup - corresponding to changing the sign of $d_{r\alpha}$ - is preserved.  As promised, the low energy theory of fluctuations about the mean field is a $Z_2$ gauge theory. As this admits a deconfined phase in $3$ spatial dimensions, our mean field ansatz describes a stable state of the original boson system. This mean field ansatz also provides a wavefunction $|\psi_b \rangle$ for our bosonic fractional topological insulator:
\begin{equation}
| \psi_b \rangle = P_{\vec S_r = 0} |\psi_{TI} \rangle
\end{equation}
Here $|\psi_{TI} \rangle$ is simply the Slater determinant wavefunction for the strong topological insulator. The operator $P_{\vec S_r = 0}$ projects onto the sector with zero pseudospin at each lattice site.

Under a time-reversal transformation $\Theta$, the boson $b$ remains invariant. How does time-reversal act on the slave fermions $d_\alpha$? The specification of
how various physical symmetries act on the slave particles is part of the formulation of the slave-particle theory; any choice is allowed as long as it is consistent with the transformation properties of physical, gauge-invariant operators. In the case at hand, it is convenient to choose $d_\alpha$ to transform as ordinary fermions: $\Theta d_{r1} = d_{r2}$ and $\Theta d_{r2} = - d_{r1}$. This choice makes the mean-field Hamiltonian of (\ref{mfHam}) manifestly time-reversal invariant and the action of $\Theta$ on the fermions satisfies $\Theta^2 = -1$. A different choice would make the mean-field Hamiltonian time-reversal invariant
only up to an $SU(2)$ gauge transformation, but would yield the same projected wave function and the same low energy physics.

The resulting low energy theory consists of four bands of fermions (two from the two atom unit cell and two from the pseudo-spin index) coupled to a $Z_2$ gauge field. The $Z_2$ gauge field is in its deconfined phase.  The choice of one boson per unit cell gives us two emergent fermions per unit cell, and these fermions form a topological insulating state by filling the lower two of the four bands.  The low energy physics is thus fully gapped in the bulk with the quasiparticles carrying charge under a $Z_2$ gauge field.  Line defects that carry $Z_2$ gauge flux (vison lines) will also exist as excitations in the bulk.  The fermionic quasiparticles carry fractional charge $1/2$ under the $U(1)$ particle number symmetry. Additionally, these line defects will carry gapless fermionic states bound to their cores \cite{1dmodes1}.

With this different minimal charge we may expect an interesting $\theta$ angle in the bulk. If the fermions carried charge $1$ under the $U(1)$ symmetry then upon integrating them out we would obtain a $\theta$ term with $\theta\, (\text{mod } 2\pi) = \pi$.  Since the fermions actually couple with fractional charge, repeating the calculation gives a $\theta$ term with $\theta\, (\text{mod } 2\pi q^2) = q^2 \pi $ where $q = 1/2$ is the fermion charge.  In fact, the $d$ fermions couple to both the $Z_2$ and the $U(1)$ gauge fields, and the result of integrating them out is a $\theta$ term for the combined gauge field $\frac{1}{2} A_{U(1)} + A_{Z_2}$ where the $Z_2$ gauge field $A_{Z_2}$ comes from the original $SU(2)$ gauge structure that was broken by the fermion band structure \cite{z2theta1,z2theta2,fti1}.

The appearance of an effectively reduced periodicity for the electromagnetic $U(1)$ $\theta$ angle is subtle.  Consider the theory with $\theta = 2 \pi q^2 = \pi/2$ on a space without boundary.  This value of the $\theta$ angle must be effectively trivial in a phase with charge $1/2$ fractionalized excitations.  The precise meaning of this statement is as follows: all physical observables on this closed space, including the spectrum of dyons and all Berry's phases, are equivalent to those of a theory with $\theta = 0$.  In the presence of a boundary there is the possibility of surface states, unprotected by time reversal, equivalent to an \textit{integer} quantum hall state of charge $1/2$ fermions, but it remains true that all bulk observables depend on $\theta \text{ mod } 2\pi q^2$.

With $q=1/2$ we find $\theta = \pi/4$ for the background $U(1)$ gauge field. One way to understand the correctness of this result is to consider
the spectrum of dyons, bound states of electric and magnetic charge. In the presence of deconfined charge $e/2$ excitations, the $2 \pi/e$ monopoles become
confined and the minimal monopole strength becomes $g_{\text{min}} = 4 \pi /e$. The Witten effect attaches charge $\frac{\theta}{2 \pi} \frac{e g_{\text{min}}}{2 \pi} e$ to the minimal monopole \cite{thetadyons}.  Here the minimal monopole carries $1/4$ extra $U(1)$ charge due to the $\theta$ term.  This extra charge is precisely half that of the minimal pure electric charge and ensures that the spectrum of dyons is symmetric under time reversal.  Another way to understand the value of $\theta$ is to consider the surface states \cite{tirev1,tirev2}. $\theta = \pi/4$ corresponds to a surface Hall conductivity of $\frac{1}{8}\frac{e^2}{2 \pi}$ \cite{tft1} which is precisely what we would obtain from a single Dirac cone of charge $1/2$ fermions.

The phase we have constructed has a gap to all excitations, a fractional $\theta$ angle, and preserves time reversal.  As we have argued, it must have ground state degeneracy on topologically non-trivial spaces such as the three torus $T_3$.  $T_3$ possesses three elementary non-contractible loops ($1$-cycles) corresponding to the $x$, $y$, and $z$ circles. By moving a $Z_2$ charge around the $i$th $1$-cycle we can detect the presence $n_i = 1$ or absence $n_i = 0$ of a $Z_2$ flux through that cycle.  Thus the ground states can be labeled by configurations $\{n_i\}$ of $Z_2$ flux through the non-contractible loops, and we find a total of $2^3 = 8$ states.

We can also consider the analog of the ground state to ground state process considered above. $2 \pi/e$ monopoles are confined in our phase by a string consisting of an odd number of $Z_2$ flux lines.  For example, it costs an energy of order the linear system size to put a background $\int_{xy} e B_z = 2 \pi$ flux because of the necessary presence of a $Z_2$ flux line wrapping the system in the $z$ direction.  However, $\int_{xy} e B_z = 4 \pi$ is a perfectly allowed low energy configuration in accord with the presence of charge $1/2$ excitations.  Similarly, inserting $\int_{t z} e E_z = 2 \pi$ flux through the $z$ cycle does not return us the same ground state as this process is equivalent to inserting a $Z_2$ flux.   Note that the phase acquired by the state in this process is not gauge invariant or physically meaningful as we have not made a closed loop in Hilbert space.  Instead, we must insert $\int_{t z} e E_z = 4 \pi$ flux to return to the same state.  The time reversed processes of inserting $4 \pi/e $ flux in a $\pm 4 \pi /e $ background flux take states to themselves up to a well defined phase.  These phases are $\exp{(\pm i 4 \theta)}$ which with $\theta/\pi = 1/4$ are equal, consistent with time reversal.
% change here

The phase we have constructed has degenerate ground states on topologically non-trivial spaces without boundary.  We can also consider interfaces between different phases: the trivial insulator $\mathcal{I}$, the $Z_2$ fractionalized $\mathcal{TI}^*$ phase we have constructed, and a more traditional $Z_2$ fractionalized phase $\mathcal{I}^*$ with $\theta = 0 $. Such a phase corresponds to choosing a non-topological insulator band structure for the $d$ fermions.  There are three kinds of interfaces we can construct from these three phases.

An $\mathcal{I}$/$\mathcal{I}^*$ interface will not generically have protected surface states, although it may admit gapless surface states in the presence of additional symmetries. An $\mathcal{I}^*$/$\mathcal{TI}^*$ interface will have protected gapless surface states just like at an interface between a fermionic topological insulator and a trivial band insulator.  The only remaining question is what happens at an $\mathcal{I}$/$\mathcal{TI}^*$ interface?  The double ``sandwich'' interface $\mathcal{I}$/$\mathcal{I}^*$/$\mathcal{TI}^*$ should have gapless surface states even as the width of the $\mathcal{I}^*$ region shrinks because the $\mathcal{I}$/$\mathcal{I}^*$ interface brings no additional gapless states to the $\mathcal{I}^*$/$\mathcal{TI}^*$ interface.  Hence we expect an $\mathcal{I}$/$\mathcal{TI}^*$ interface to have protected surface states.

The conclusion that we have gapless surface states is supported by the presence of the bulk $\theta$ angle. Indeed, if the surface of our $\mathcal{TI}^*$ phase is covered with a time-reversal breaking perturbation then the bulk $\theta$ term implies that the surface Hall conductivity is $\sigma_{xy} = (2n+1)\frac{1}{8} \frac{e^2}{2\pi} $.  This is effectively the contribution from a single Dirac cone, protected in the absence of time reversal breaking, modulo integer quantum hall states of the charge $1/2$ fermions.

One can generalize our construction to provide examples of fractionalized fermionic topological insulators. For example, first fractionalize the physical spinful fermion $c_{r\alpha}$ into a fermionic spinon $f_{r\alpha}$ and a bosonic chargon $b_r$ so that  $c_{r\alpha} = b_r f_{r\alpha}$ in a $Z_2$ deconfined phase while preserving all symmetries.  Then put the boson $b_r$ into an interesting state of the type described here with fractionalized charge-$1/2$ fermions that have a topological band structure.  This will yield a stable fractional topological insulator of fermions with $\theta = \frac{\pi}{4}$ and will preserve time reversal symmetry.

The possibility of time reversal invariant fractional topological insulating phases for fermions was posed very recently in \cite{fti1} that partially motivated this work. The specific theoretical construction of such a phase in that work also involved a slave particle construction that breaks the electron into $N$ fractionally charged `quarks' with an $SU(N)$ gauge redundancy.  Each of the $N$ quarks were assumed to form strong topological insulators,
and upon integrating out these quarks, the low energy theory was argued to be $SU(N)$ gauge theory with $\theta_{SU(N)} = \pi$ and $\theta_{U(1)} = \pi/N$. As that work points out, the flow to strong coupling is believed to be confining for these quarks. Thus the mean field quark topological band structure ansatz does not, in that case, directly describe a stable phase. What is the precise low energy fate once the $SU(N)$ gauge theory flows to strong coupling?   There is considerable uncertainty in answering this question due to our rather poor understanding of strongly coupled non-abelian gauge theories. In the large $N$ limit \cite{thetagauge} it is known that $SU(N)$ gauge theory at $\theta_{SU(N)} = \pi$ spontaneously breaks time reversal. Thus at least at large $N$ this construction does not lead to a time reversal protected topological insulator.  The situation is less clear at small $N$.  However, if time-reversal symmetry is not spontaneously broken and there is a bulk energy gap, then our argument requires that the theory possess ground state degeneracy on $T_3$ and fractionalized excitations.  One way this may happen is if the theory enters a Higgs phase that leaves unbroken the $Z_N$ center of $SU(N)$, similar to our construction.

We have studied a number of questions surrounding correlated topological insulators in interacting bosonic and fermionic systems. Such insulators admit the possibility of having fractional $\theta/\pi$ while preserving time reversal, as suggested in \cite{fti1}. We showed that this requires the presence of deconfined fractional excitations in the bulk and associated ground state degeneracies on topologically non-trivial closed spaces. We also showed that even repulsive bosons could form topological insulators with time reversal protected surface modes.  This was illustrated with a concrete example of a time
reversal invariant $Z_2$ fractionalized phase of bosons with fractional $\theta/\pi = 1/4$, degenerate ground states on the three-torus $T_3$, and protected surface states.

While our results show that fractional topological insulators are necessarily fractionalized in the more general sense, the converse is of course not true: fractionalized phases need not have fractional $\theta/\pi$.  We also conjecture that there are no non-fractionalized phases of bosons with $\theta = \pi$ based on the intuition that one must always have fermionic excitations in the spectrum to achieve a topological insulating state. This remains an open question.

There are a number of directions for future work.  It would be interesting to identify some candidate materials where bosonic topological insulators might be realized. To that end it might be useful to reinterpret them as possible phases of insulating spin-$1/2$ quantum magnets with conserved $S^z$. In that incarnation they correspond to gapped $Z_2$ quantum spin liquids with protected gapless surface modes. More generally, correlated phases with interesting protected surface modes are likely to support a rich set of phenomena that will hopefully be explored in detail in the future.

We thank M. Levin, X.-L. Qi, J. Maciejko, A. Karch, and T. Grover for useful discussions.  TS was supported by NSF Grant DMR-0705255. JM was supported by the U.S. Department of Energy under cooperative research agreement DE-FG0205ER41360 and by an Alfred P. Sloan fellowship.

\bibliography{fti}

\end{document}